\documentclass[twoside,english,12pt]{cern2}

\usepackage{amssymb}

\usepackage{graphicx}
\usepackage{babel}
\usepackage{multicol}
\usepackage[centerlast]{caption2}
\usepackage[section]{placeins}

\newcommand{\registered}{\circledR}

\begin{document}

\title{Secondary Electron Yield Measurements of TiN Coating and TiZrV Getter Film}
\author{F. Le Pimpec, F. King, R.E. Kirby, M. Pivi \\
SLAC, 2575 Sand Hill Road, Menlo Park, CA 94025 }
\date{9th October 2003}
\maketitle

\thispagestyle{headings}
\markright{SLAC-TN-03-052 $\backslash$ revised 27th August 2004}

\begin{abstract}
In the beam pipe of the positron Main Damping Ring (MDR) of the
Next Linear Collider (NLC), ionization of residual gases and
secondary electron emission give rise to an electron cloud which
can cause the loss of the circulating beam. One path to avoid the
electron cloud is to ensure that the vacuum wall has low secondary
emission yield and, therefore, we need to know the secondary
emission yield (SEY) for candidate wall coatings. We report on SEY
measurements at SLAC on titanium nitride (TiN) and
titanium-zirconium-vanadium (TiZrV) thin sputter-deposited films,
as well as describe our experimental setup.

\end{abstract}

\section{Introduction}

Beam-induced multipacting, which is driven by the electric field
of successive positively charged bunches, arises from a resonant
motion of electrons that were initially generated or by gas
ionization or by secondary emission from the vacuum wall. These
electrons then bounce back and forth between opposite walls of the
vacuum chamber. The electron cloud density depends on
characteristics of the positively charged circulating beam (bunch
length, charge and spacing) and the secondary electron yield and
spectrum of the wall surface from which the starting electrons
arise. The electron cloud effect (ECE), due to multipacting, has
been observed or is expected at many storage rings
\cite{Pivi:pac03}. The space charge of the cloud, if sufficient,
can lead to a loss of the beam or, at least, to a drastic
reduction in bunch luminosity.

\begin{figure}[tbph]
\begin{center}
\includegraphics[width=0.8\textwidth,clip=]{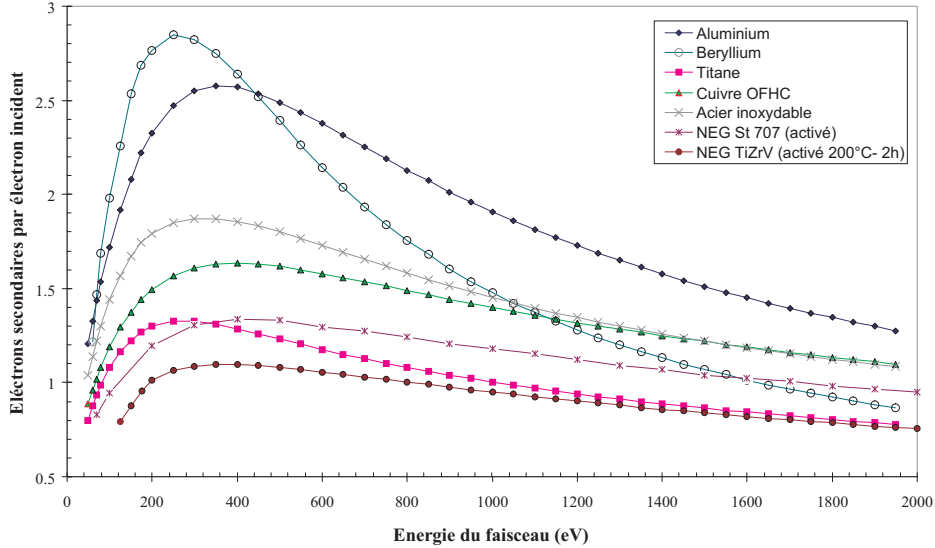}
\end{center}
\caption{SEY of baked technical surfaces. 350$^\circ$C for 24hr
\cite{flp}} \label{figSEYflpphd}
\end{figure}

\medskip
In order to minimize the electron cloud problem which might arise
in the NLC, we are looking to a solution involving surface coating
of the secondary electron emitting vacuum wall. The SEY of
technical surfaces has been measured in the past at SLAC
\cite{kirby:1995} \cite{kirby:2001}, at CERN
Fig.\ref{figSEYflpphd} \cite{Hilleret} \cite{Hilleret:EPAC00} and
in other labs \cite{Kato:AVS2002}. In this paper we present
measurements of the SEY of materials previously measured and known
to have low SEY \cite{laurent:0304} \cite{Scheurlein:2002}
\cite{he:PAC01} like titanium-nitride thin film (TiN) and
titanium-zirconium-vanadium getter film (TiZrV).

\section{Experiment Description}
\label{expsetupdescript}

\begin{figure}[tbph]
\begin{center}
\includegraphics[width=0.85\textwidth,clip=]{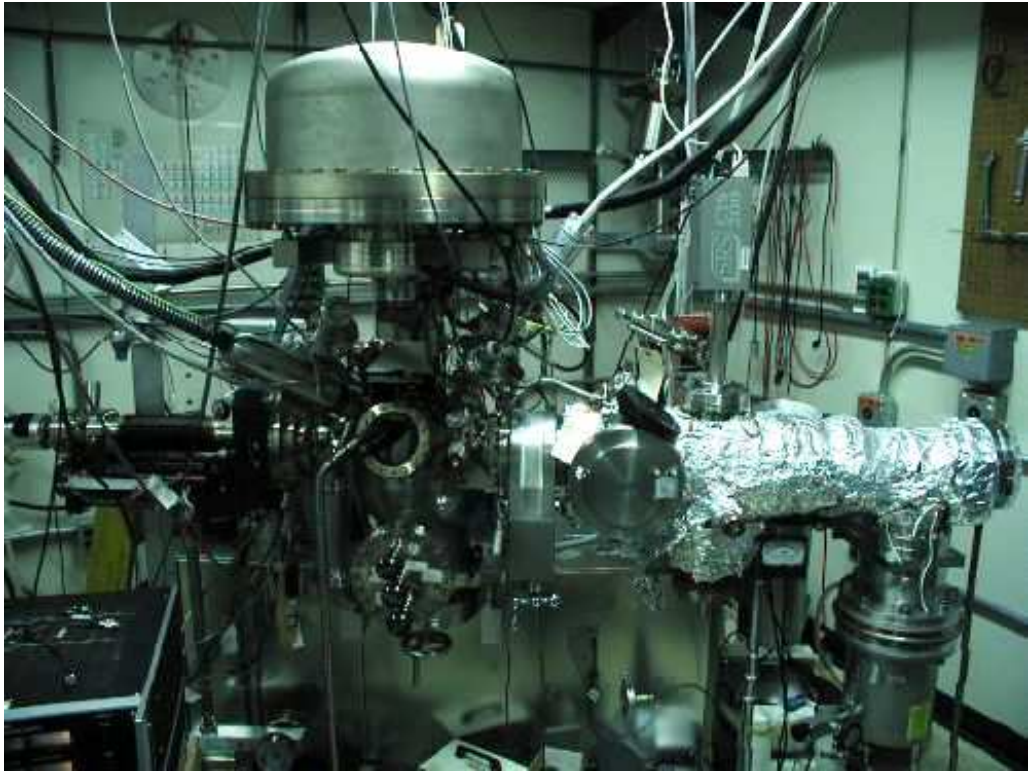}
\end{center}
\caption{Experimental system used for surface analysis}
\label{figSEYkirbysetup}
\end{figure}

The system used to measure SEY is composed of two coupled
stainless steel (S/S) UHV chambers where the pressure is in the
low 10$^{-10}$ Torr scale in the measurement chamber and high
10$^{-9}$ Torr scale in the "load lock" chamber,
Fig.\ref{figSEYkirbysetup}. Samples individually screwed to a
carrier plate, are loaded first onto an aluminium transfer plate
in the load lock chamber, evacuated to a low 10$^{-8}$ Torr scale,
and then transferred to the measurement chamber.

\medskip
The measurement chamber has two electron guns and a soft
(1.49~keV) x-ray source. One electron gun (energy, 1-20~keV) is
used for Auger electron spectroscopy (AES) light element surface
contamination analysis. The x-ray source is used to excite
photoelectrons for surface chemical valence analysis, called ESCA
(Electron Spectroscopy for Chemical Analysis). TiN stoichiometry
is measured by ESCA technique which is also called XPS (X-ray
Photoelectron Spectroscopy).

\medskip
The principle of XPS is to collect photoelectrons ejected by
x-rays of known energy near the surface (1 - 5~nm information
depth). The emitted electrons have an energy E$_k$ which is given
by equation~\ref{equXPS}

\begin{equation}
E_k = h\nu - E_b -\Phi
\label{equXPS}
\end{equation}

where h$\nu$ is the energy of the incident photon, E$_b$ the
binding energy of the electron relative to the Fermi level of the
material and $\Phi$ the spectrometer work function. The spectrum
of the measured kinetic energy gives the spectrum of the binding
energy of the photoelectrons.

\medskip
The x-ray source is also used for exciting secondary X-ray
Fluorescence (XRF) for thickness measurement of the deposited TiN
overlayers. The second electron gun (0-3~keV) is used measure the
SEY, and can also be used to electron condition the
surface. An ion gun is available for cleaning the samples by
sputtering and for ion conditioning surfaces.

\begin{figure}[tbph]
\begin{center}
\includegraphics[angle=-90,width=0.75\textwidth,clip=]{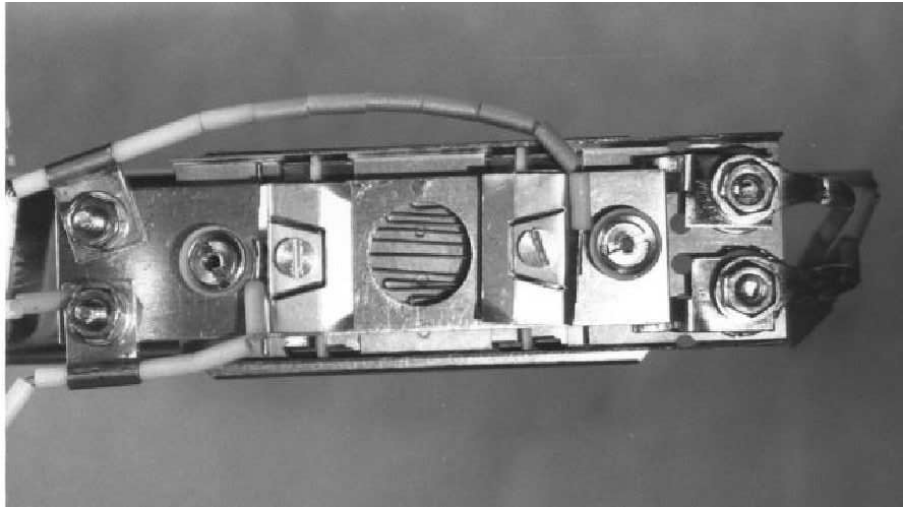}
\end{center}
\caption{Omniax sample holder plate with heating filament visible}
\label{figOmniaxholder}
\end{figure}

\begin{figure}[tbph]
\begin{center}
\includegraphics[width=0.75\textwidth,clip=]{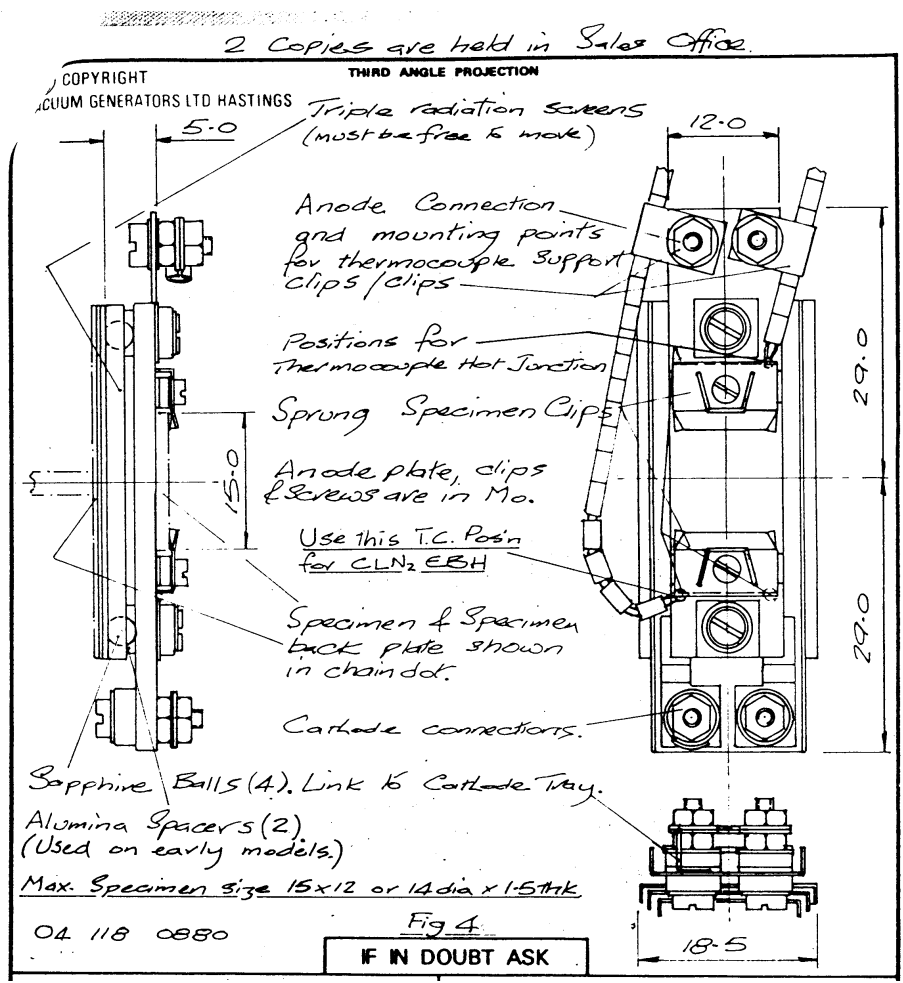}
\end{center}
\caption{Drawing of the Omniax$^{\registered}$ Sample Holder plate
without the heating filament. \copyright Vacuum Generator Ltd}
\label{figOmniaxholderDrwg}
\end{figure}

\medskip
After all samples (up to ten or so) are transferred into the
measurement chamber, one sample at a time is loaded, on its
individual carrier plate, onto a manipulator arm (Vacuum
Generators "Omniax"). The Omniax$^{\registered}$ carrier plate
holder design is shown in Fig.\ref{figOmniaxholderDrwg}. Two
thermocouples are installed on this holder plate as well as
heating filament and sample connection wires. Samples can be
heated via a tungsten wire filament, Fig.\ref{figOmniaxholder}, by
radiation or by electron bombardment. Electron bombardment is
achieved by biasing the filament negatively. The Omniax sample
holder is insulated from ground via an alumina ceramic and by
several thermal shield plates insulated via four sapphire balls,
Fig.\ref{figOmniaxholder} and Fig.\ref{figOmniaxholderDrwg}.


\begin{figure}[tbph]
\begin{center}
\includegraphics[width=0.8\textwidth,clip=]{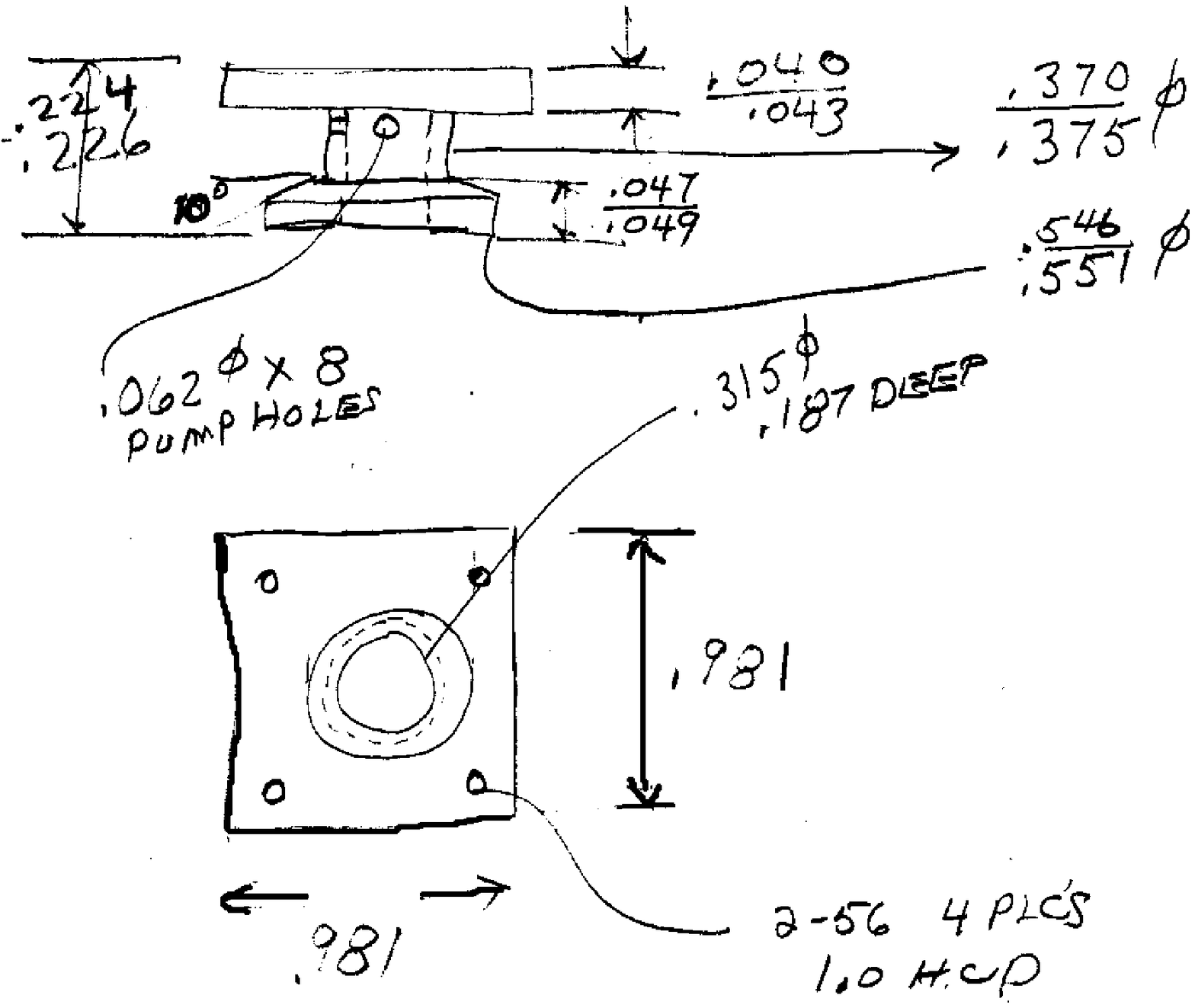}
\end{center}
\caption{Sketch of the Sample carrier plate made in stainless
steel. Dimensions are in inches }
\label{figSampleHolderDrwg}
\end{figure}

\begin{figure}[tbp]
\begin{minipage}[t]{.5\linewidth}
\centering
\includegraphics[width=0.9\textwidth,clip=]{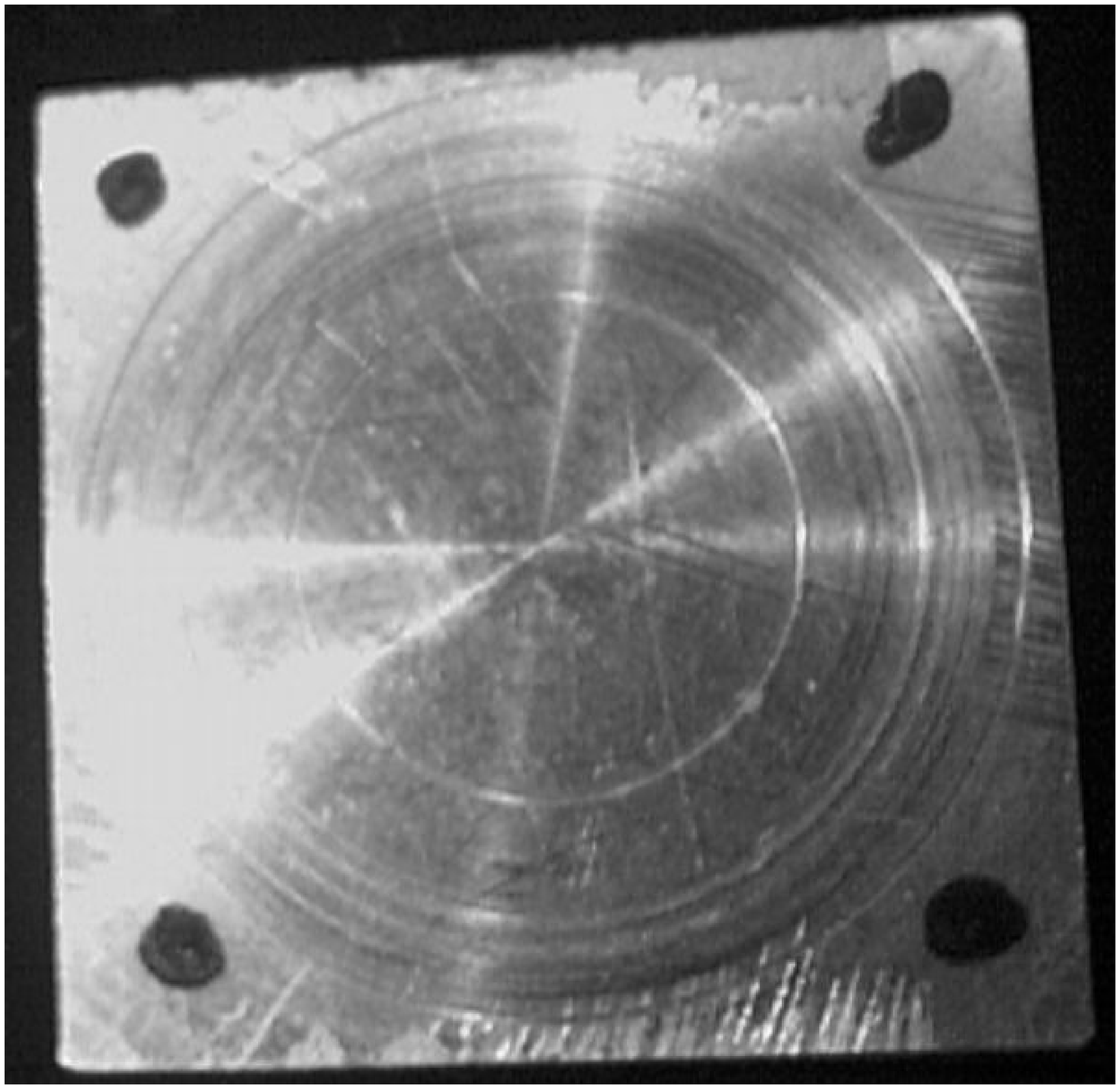}
\setcaptionwidth{6cm} \caption{Sample holder prototype in
aluminium. The black dots are the location of the screw sample
mounting holes}
\label{figSampleHolderface}
\end{minipage}%
\begin{minipage}[t]{.5\linewidth}
\centering
\includegraphics[width=0.9\textwidth,clip=]{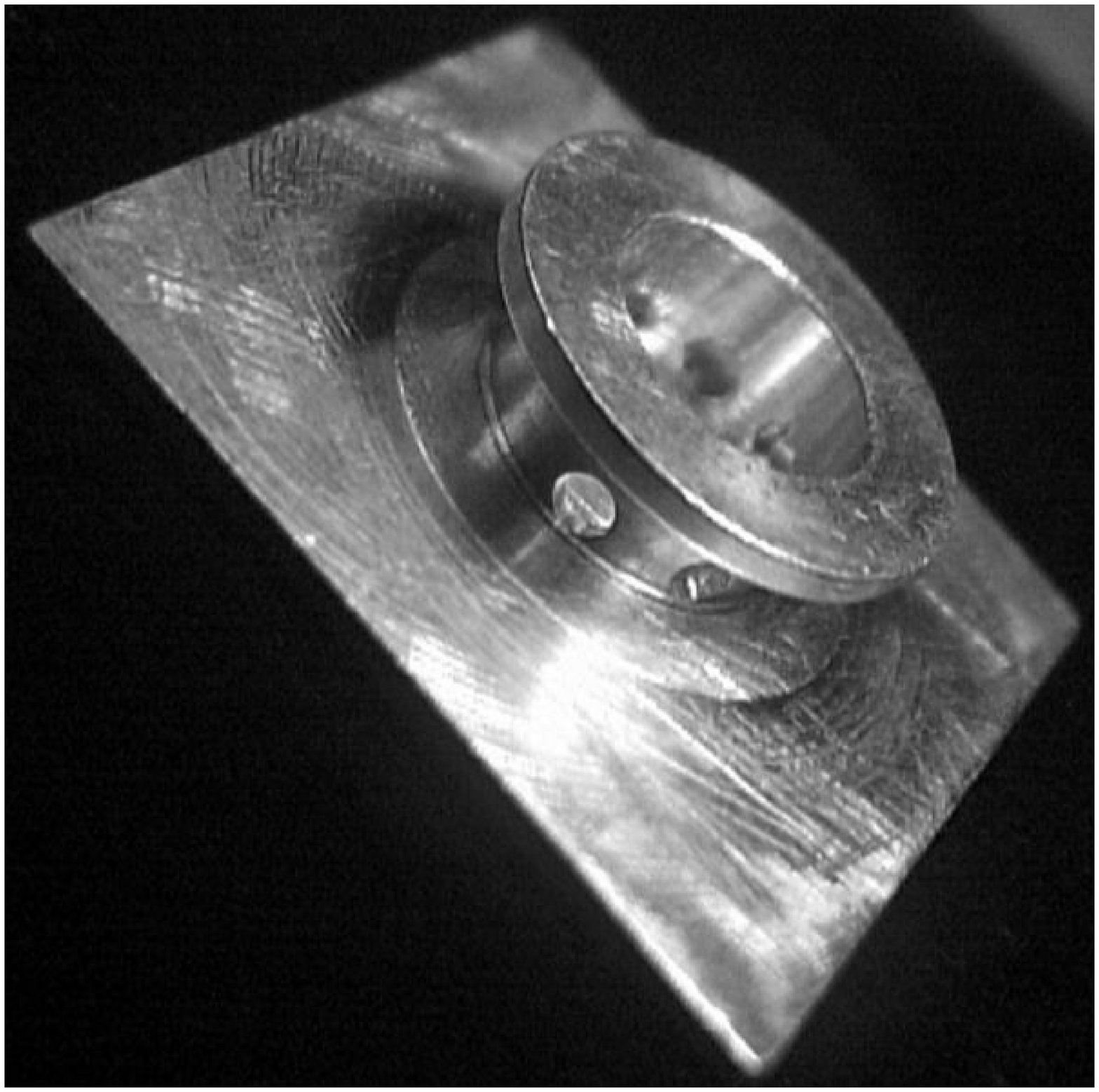}
\setcaptionwidth{6cm}
\caption{Bottom side of the sample holder}
\label{figSampleHolderbck}
\end{minipage}
\end{figure}

\medskip
The sample carrier plates are made of molybdenum (for attaining
the highest temperatures) or stainless steel,
Fig.\ref{figSampleHolderDrwg}. This plate will support one sample
which is held by corner screws onto its surface. This design
allows the plate to then slide into the rails of the Omniax plate
holder, cf Fig.\ref{figOmniaxholder}. An example of an aluminium
prototype carrier plate is shown in Fig.\ref{figSampleHolderface}
and Fig.\ref{figSampleHolderbck}. Note that on
Fig.\ref{figSampleHolderface} the holes for the screws are not
drilled. In this configuration, the thermocouples are not attached
directly to the carrier plate but indirectly through the holder
plate. Sample temperatures are compared to the thermocouple
temperatures by using a black body-calibrated infrared pyrometer
(0.8~$\mu$m - 1.3~$\mu$m bandpass) which is corrected for
absorption by the 7056 glass viewport of the measurement chamber
and the sample emissivity ($\epsilon$) of 0.2~-~0.4.

\medskip
A good way to monitor the activation process of the TiZrV
non-evaporable getter (NEG) is to record the decrease of the
surface oxygen concentration with XPS. During the NEG activation,
the surface goes from an oxidized state to a partially metallic
state. The backside of the NEG was heated, directly via a hole in
the plate holder, by electron bombardment. The minimum temperature
needed to activate this NEG is 180$^\circ$C \cite{Scheurlein:2002}
\cite{benven:2001}. The thermocouples displayed a temperature of
216$^\circ$C for 2~hours. In order to determine the temperature of
the NEG, a S/S sample holder ($\epsilon\simeq$ 0.31) was heated,
with a hot plate, to various temperature with a thermocouple spot
welded to it, with and without 7056 glass viewport correction.
Matching the reading of the pyrometer of this experimental
calibration with the reading recorded during an in situ heating of
the same S/S sample holder allowed us to determine the actual
temperature of the NEG, assuming the same emissivity,
T$_{NEG}$~=~232$^\circ$C. An in-situ temperature measurement with
the black body calibrated pyrometer gives for the same S/S surface
201$^\circ$C when the thermocouples on the Omniax carrier plate
holder read 216$^\circ$C.

\begin{figure}[tbph]
\begin{center}
\includegraphics[width=0.45\textwidth,clip=]{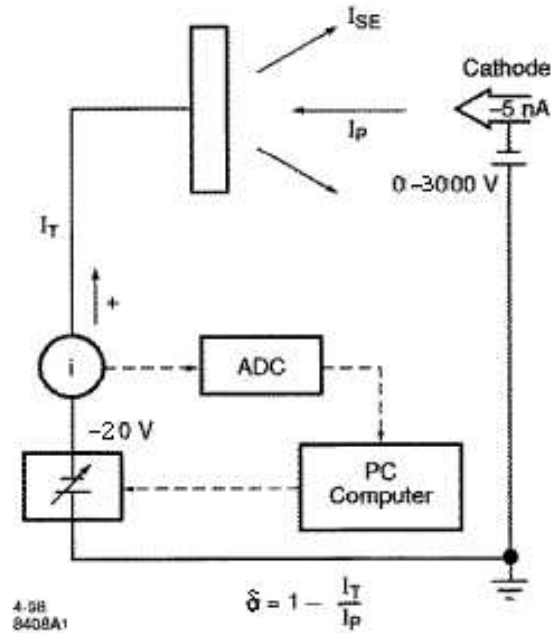}
\end{center}
\caption{Electronic circuitry used to measure the secondary emission yield}
\label{figelctrqcircuit}
\end{figure}

\medskip
The experimental electronic circuit is identical to the one
presented in Fig. \ref{figelctrqcircuit} \cite{kirby:2001}. The
computer controlled electron beam coming from the gun is decoupled
from the target measurement circuitry. However, the ground is
common to both. In short, the target is attached to a biasing
voltage source and an electrometer connected in series to the data
gathering computer Analog Digital Converter (ADC).

\section{SEY measurement methodology}

\begin{figure}[tbph]
\begin{center}
\includegraphics[width=0.85\textwidth,clip=]{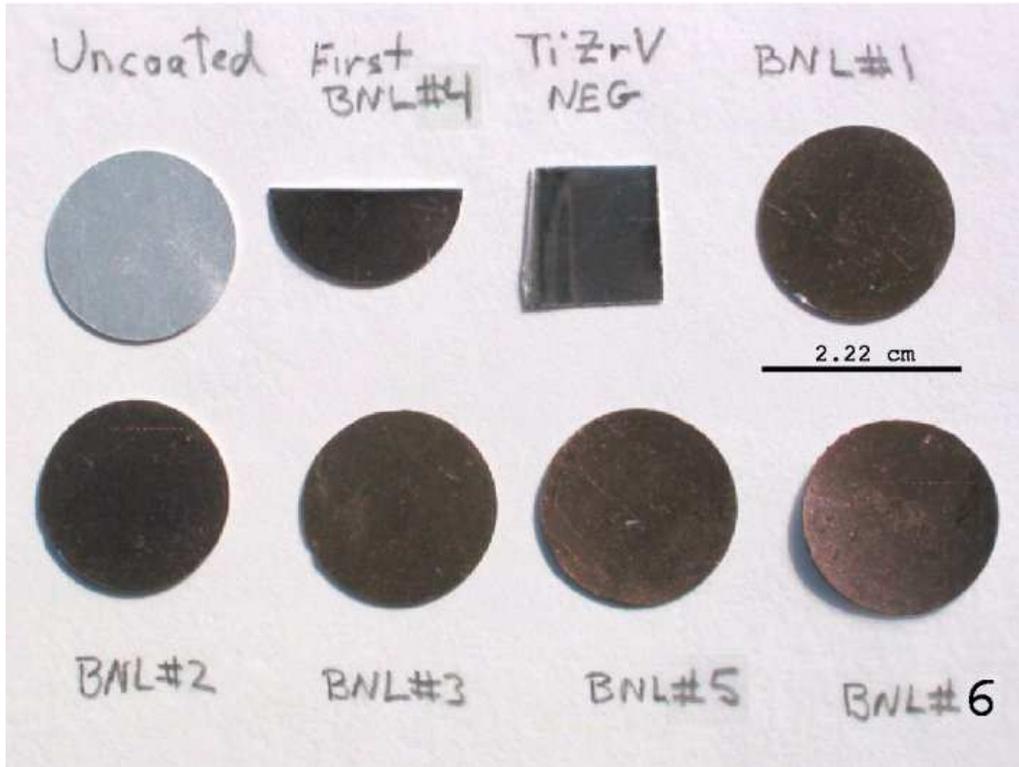}
\end{center}
\caption{Samples measured thus far.}
\label{figSEYSamplesBatchOne}
\end{figure}

In order to characterize our system, we have measured the SEY
in-situ of inert gas-ion cleaned materials: carbon, copper and
gold. Results are not presented here but agree with widely
published results for ion sputter cleaned sample of these
materials. After confirming the SEY from these reference
materials, we proceeded to measure a TiN-coated aluminium samples
provided by BNL and a TiZrV sputter deposited film on stainless
steel substrate, obtained from CERN,
Fig.\ref{figSEYSamplesBatchOne}.

SEY ($\delta$) definition is given in
equation~\ref{equdefinition}. In practice equation~\ref{equSEY} is
used because it contains parameters measured directly in the
experiment.

\begin{minipage}[h]{.6\linewidth}
\begin{equation}
\delta = \frac{Number\ of\ electrons\ leaving\ the\
surface}{Number\ of\ incident\ electrons}
\label{equdefinition}
\end{equation}
\end{minipage}
\begin{minipage}[h]{.35\linewidth}
\begin{equation}
\delta = 1 -\frac{I_T}{I_P}
\label{equSEY}
\end{equation}
\end{minipage}

\medskip
Where I$_P$ is the primary current or the current leaving the
electron gun and impinging on the surface of the sample and I$_T$
is the total current measured on the sample ($I_T = I_P + I_S$).
I$_S$ is the secondary electron current leaving the target.

\begin{figure}[tbp]
\begin{minipage}[t]{.5\linewidth}
\centering
\includegraphics[width=0.9\textwidth,clip=]{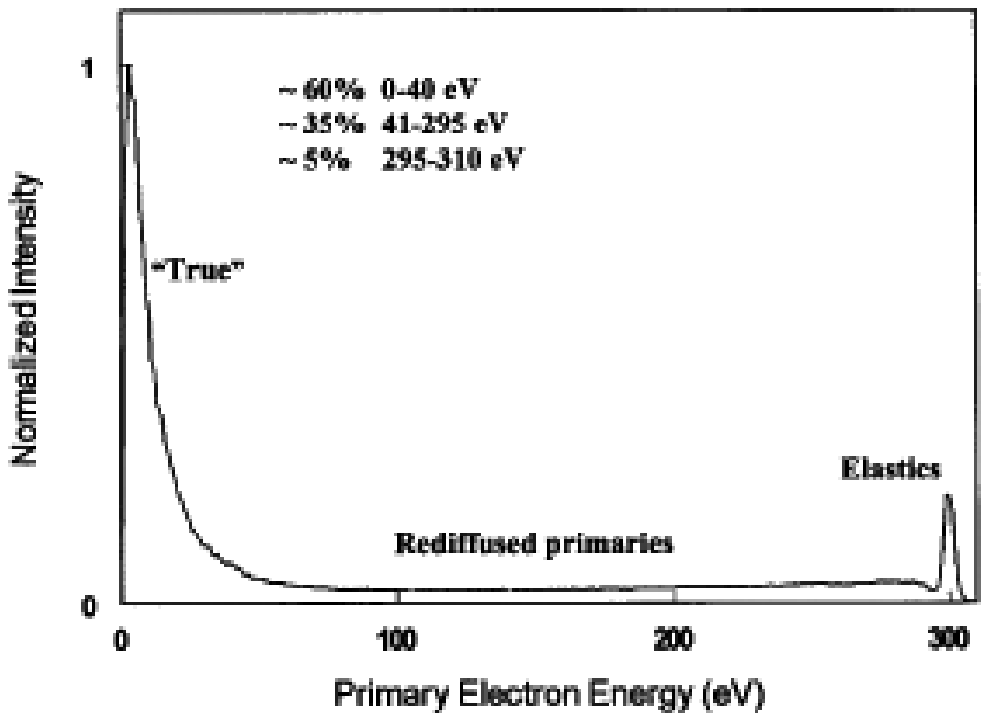}
\setcaptionwidth{6cm}
\caption{Spectrum of a secondary electron beam from a 300~eV
incident primary beam impinging on a TiN on Al substrate sample
\cite{kirby:2000}}
\label{figspctrumelect}
\end{minipage}%
\begin{minipage}[t]{.5\linewidth}
\centering
\includegraphics[width=0.92\textwidth,clip=]{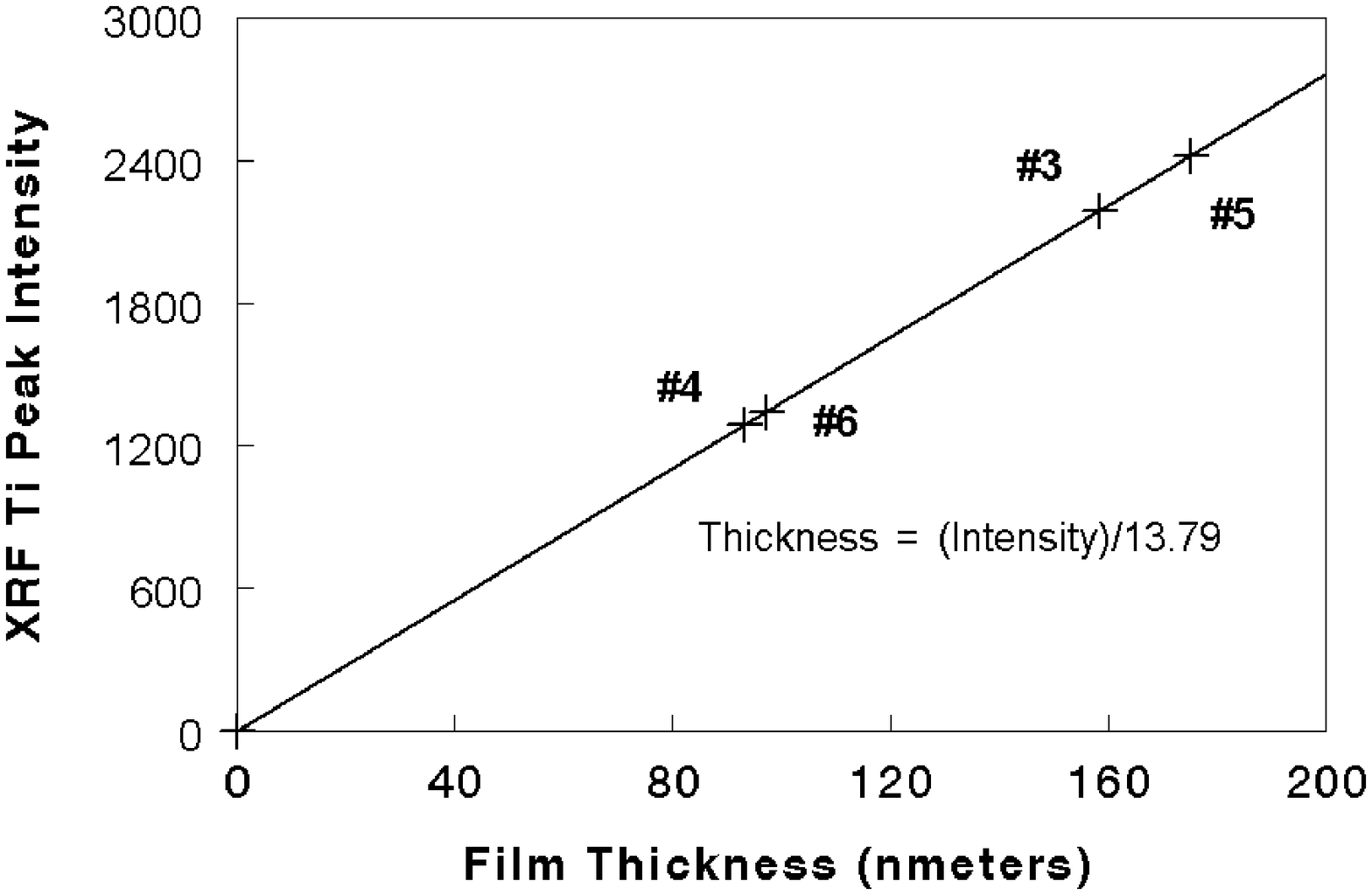}
\setcaptionwidth{6cm}
\caption{TiN thickness measured by x-ray
fluorescence of the Ti K-line of four BNL samples}
\label{figTiNthickness}
\end{minipage}
\end{figure}

\medskip
The spectrum of secondary electron current leaving the target is
composed of true secondaries (0~eV to 40~eV, by convention),
re-diffused primary electrons exiting after suffering losses in
the sample (40~eV to Ep) and from incident primary electrons (Ep)
elastically reflected from the surface, Fig.\ref{figspctrumelect}.
The majority of the electrons leaving the surface are true
secondaries.

\medskip
In order to measure the primary current leaving the electron gun,
the sample is biased at +150~V. The bias voltage prevents all
re-diffused and secondary electrons of less than 150~eV from
leaving the sample.  Elastically reflected electrons are not
collected and could strike nearby surfaces, creating secondary
electrons that are then collected by the sample bias. This effect
is small because the reflectivity at 100~eV is a few per cent. We
estimate the error in primary beam current measurement to be small
(1-2\%) because the loss of elastics is balanced by the gain of
nearby secondaries. The SEY around 100~eV for baked stainless
steel is close to 1.1, Fig.\ref{figSEYflpphd}. With regard to the
gun current as a function of energy, it starts at zero for zero
energy and smoothly increased to its saturation value at 70~eV. We
measure the magnitude and functional dependence of the beam
current up to somewhat higher value(100~eV) and use a constructed
lookup table of the beam current for SEY calculations. Not biasing
the sample with a high enough voltage will lead to an
underestimation of the beam current. This is easily understood
from the secondary spectrum, Fig.\ref{figspctrumelect}. The
selected 2~nA gun current is measured for a gun energy of 0-100~eV
by energy steps of 10~eV (0-3000~eV range) or 2~eV (0-300~eV
range).

\medskip
Actual measurement of the SEY is done by biasing the sample to
-20~V. This retarding field repels most secondaries from adjacent
parts of the system that are excited by the elastically reflected
primary beam. SEY measurements are done twice, once between 0~eV
to 3000~eV with 10~eV steps, then between 0~eV to 300~eV, with
2~eV steps. The final energy of the primary electrons is
respectively 2980~eV and 280~eV because of the bias. The primary
beam current function is measured and recorded each time before an
SEY measurement, and with the same step in energy for the electron
beam. A fresh current lookup table is created with each
measurement.
\newline The purpose of the second measurement, 0~eV to 300~eV, is to try to
understand the structure of the SEY curve at very low energy.
Several points are important, though.

\begin{enumerate}

\item Because of the negative sample bias, primary electrons near
0~eV at the sample are assured to be leaving the gun (20~eV
departure) and arriving at (20~eV -20~V)~eV.

\item Because of the algorithm used to calculate the SEY from the
primary and sample currents at 0~eV incident energy, a divide by 0
blowup occurs. To avoid this problem, the first point at 0~eV is
forced to value one. The first "true" data point is at 2~eV for
300~eV range and 10~eV for 3000~eV range.

\item The uncertainty in the sample electrometer current reading
is set by the input operational amplifier bias leakage,
$\pm$~20~pA.

\item Using sample current to determine SEY excludes the
elastically reflected electrons from the calculation (they would
need to be collected by an external to the sample grid structure).
That serves to increase the SEY (less sample current) by 1-3\%.
This is fortuitously balanced by the fact that the -20~V bias does
not repel 100\% of nearby surface secondaries.

\end{enumerate}

The consequence of all of these points is that the SEY
measurements are just that, "secondary". Which means that it does
not include the elastics. The SEY measurements are however,
accurate. It is important to not look at the SEY at low primary
energy and try to conclude something about elastic reflectivity.
Data below 20~eV comes from a band structure and are a combination
of diffraction from the crystalline structure and energy
absorption by the material \cite{kirby:1985}. Surface effects such
as roughness can also change the SEY.

\section{Results and Comments \label{results}}

The TiN coating, made at Brookhaven National Laboratory (BNL), was
deposited onto aluminium alloy substrates, following the same
recipe described in \cite{he:PAC01}. For the Spallation Neutron
Source (SNS) project, the coating was done on S/S. According to
BNL, the expected film thickness is around 1000~\AA. We measured
the actual sample thicknesses using XRF,
Fig.\ref{figTiNthickness}.

\medskip
The principle of XRF is to collect secondary x-rays generated and
exiting the sample when bombarded by primary incident x-rays. The
secondary fluorescence yield is highest somewhat above the
K-absorption edge of Ti, so primary x-rays of 7~keV are used for
the excitation of the Ti-K$_{\alpha}$ line (4.51~keV). A TiN film
of similar known thickness (by Rutherford backscatter
spectrometry, performed at an outside lab) is used to calibrate
the technique.\newline The height of the measured Ti-K$_{\alpha}$
line, using a Si Li-drifted x-ray detector, is linearly
proportional to the number of Ti atoms in the film. Results are
shown for a few samples in Fig.\ref{figTiNthickness}.

\begin{figure}[htbp]
\begin{minipage}[t]{.5\linewidth}
\centering
\includegraphics[width=0.85\textwidth,clip=]{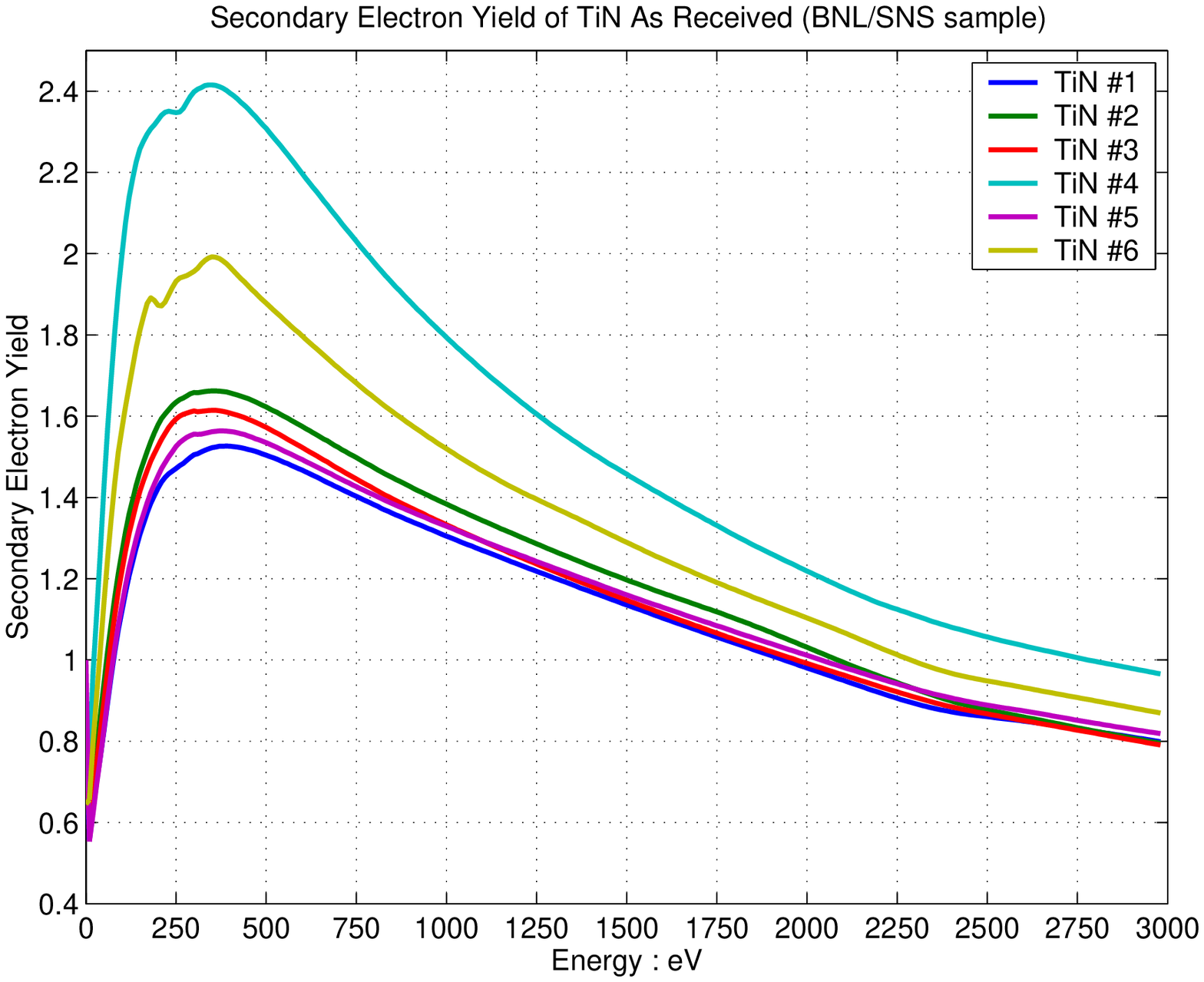}
\setcaptionwidth{6cm}
\caption{SEY of different TiN sample for electron energy between 0-2980~eV}
\label{figTiN0-3000ev}
\end{minipage}%
\begin{minipage}[t]{.5\linewidth}
\centering
\includegraphics[width=0.85\textwidth,clip=]{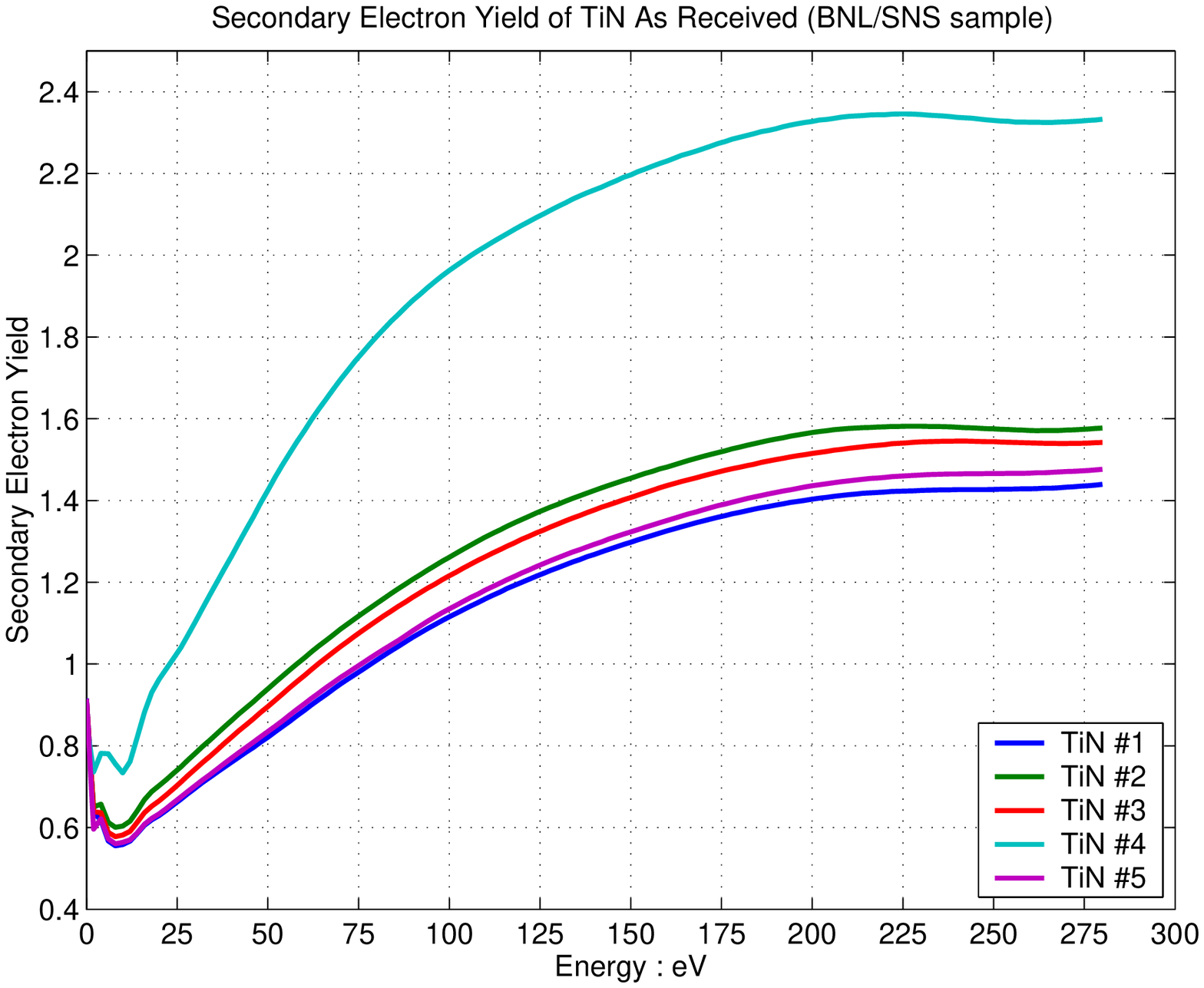}
\setcaptionwidth{6cm}
\caption{SEY of different TiN sample for electron energy between 0-280~eV.}
\label{figTiN0-280ev}
\end{minipage}
\end{figure}

\medskip
SEY measurements results of six different "as received" TiN
samples are displayed in Fig.\ref{figTiN0-3000ev} and
Fig.\ref{figTiN0-280ev}. The electron beam impinging onto the
surface is of the order of 2~nA over an area of less than a
mm$^2$. Typically the beam size is between 0.2~mm to 0.4~mm in
diameter. The low current is necessary in order to avoid surface
conditioning during SEY measurement. The size of the beam can be
checked by using a fluorescent screen, or is inferred from
secondary electron microscopical imaging (available on the
measurement system and used to precisely choose the point of SEY
measurement).

\begin{figure}[htbp]
\begin{minipage}[t]{.5\linewidth}
\centering
\includegraphics[width=0.85\textwidth,clip=]{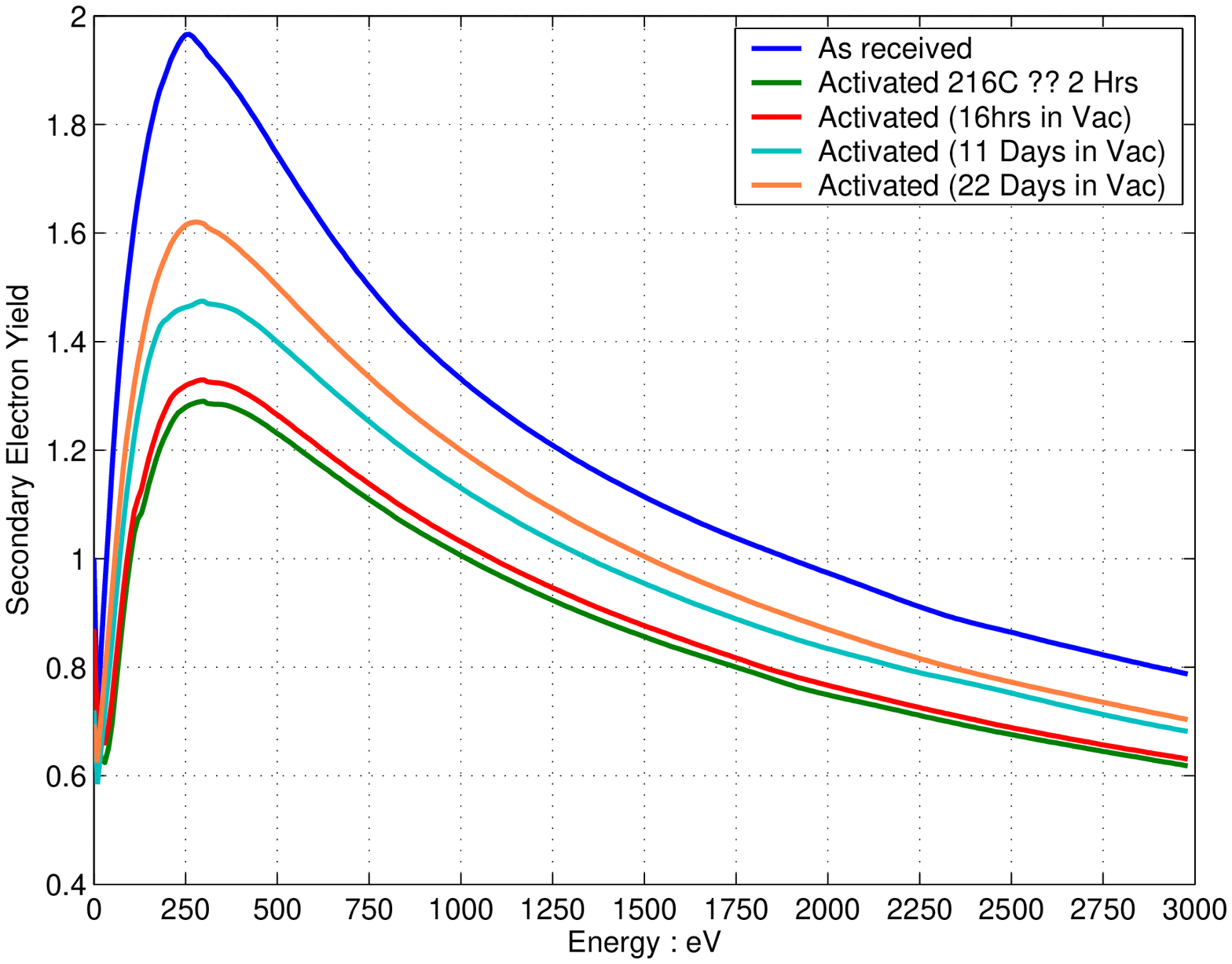}
\setcaptionwidth{6cm}
\caption{SEY of TiZrV after different process, electron energy between 0-2980~eV}
\label{figTiZrV0-3000evA}
\end{minipage}%
\begin{minipage}[t]{.5\linewidth}
\centering
\includegraphics[width=0.85\textwidth,clip=]{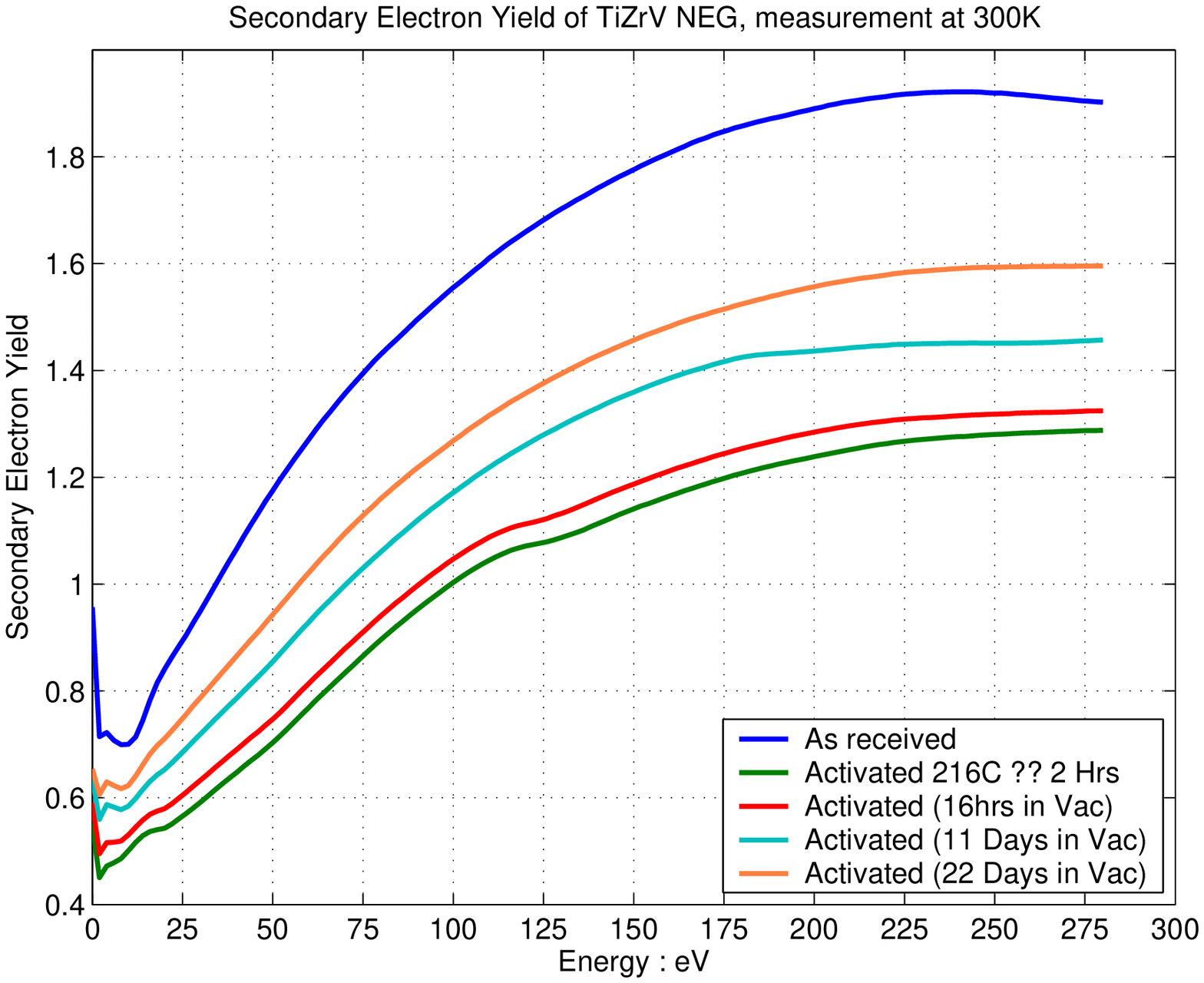}
\setcaptionwidth{6cm}
\caption{SEY of TiZrV after different process, electron energy between 0-280~eV}
\label{figTiZrV0-280evA}
\end{minipage}
\end{figure}

\medskip
The SEY of the samples varies from 1.5 to 2.5, with the thickest
film samples displaying the lower SEY. However, we have no data
concerning the roughness, and the roughness can be a factor which
can change the SEY. The commonly accepted hypothesis is that, for
a given chemical surface, the rougher surface has a lower SEY than
a smoother one \cite{Hilleret}. The irregularity in the SEY, at
near maximum, for sample TiN\#4 and TiN\#6 can be due to a non
uniform spot emitting secondaries with two different yields. The
result could be the superposition of the two SEY curves. As the
gun appears very reliable in repetitive measurements we do not
believe the irregularity is an artifact due to the system. Other
measurements of the SEY for unbaked S/S ($\delta \simeq 2$ at
250~eV) or an as received NEG, Fig.\ref{figTiZrV0-3000evA}, do not
shows these shoulders. XPS survey of TiN\#6 shows presence of
magnesium. Its expected influence on the SEY is not known.

\medskip
Finally, an alternative coating to TiN is sputter-deposited TiZrV
getter ($\sim 2 \mu$m thickness). This NEG, when activated, shows
a drastic reduction of its SEY, Fig.\ref{figTiZrV0-3000evA} and
Fig.\ref{figTiZrV0-280evA}. These results confirm what has been
investigated elsewhere \cite{Scheurlein:2002}. It is also
interesting to follow the behaviour of the SEY curves when the
sample is just exposed to only a residual gas background of
$\sim$3.10$^{-10}$~Torr for an extended period of time. The SEY of
the TiZrV goes up with time when exposed to even such good vacuum.
Interestingly enough, it is claimed in \cite{Scheurlein:2002} that
the influence in the SEY of the NEG after exposure to 30~000~L (1L=
10$^{-6}$~Torr.s) of CO or CO$_2$ is rather small. $\delta_{max}$
will increase from 1.1 (CERN fully activated NEG) to 1.35 (max)
while in UHV \cite{Scheurlein:2002}. $\Delta$(SEY) is 0.25 using
CERN results and is 0.3 for ours, $1.3 \leq \delta_{max} \leq
1.6$, Fig.\ref{figTiZrV0-3000evA}.

\medskip
If we consider, at first, that our SEY measurement is errorless,
one can argue that our activation was not complete. Hence, some
initial air formed oxide is still present and the surface
chemistry is not identical to a fully activated NEG, hence
explaining our $\delta$ after activation of 1.3. In this case the
SEY should still not increase above 1.35 \cite{Scheurlein:2002} as
the getter does not have many pumping sites left. However, this
hypothesis might have to be discarded as the minimal temperature
we are certain to have achieved is at least 201$^\circ$C.
\newline As a result of a power outage of an hour, during the eleven days of
monitoring, an ion gauge, a residual gas analyser and the ion pump
were turned off. At the switch on of the gauges, hot filaments
release gases in the system before being pumped away by the ion
pump. The recorded pressure in the system was at the switch on of
the gauges $\sim$6.10$^{-10}$~Torr. This kind of incident also can
happen in an accelerator, and it is interesting to see that this
effect leads to the recontamination of the getter. Between the
days eleven and days twenty, the system was used to XPS other
samples. Due to the transfer of samples from the load lock chamber
to the measurement chamber, the pressure rose up momentarily to
$\sim$2.10$^{-9}$~Torr.
\newline If CO or CO$_2$ exposure do not seem to affect the
increase of $\delta$, air exposure does \cite{Scheurlein:2002}. It
is possible that opening our baked measurement chamber (P~$\sim
3.10^{-10}$~Torr) to the load lock chamber (P~$\sim
9.10^{-9}$~Torr), which is frequently open to air, can be consider
as an air exposure. Hence, there is no contradiction between CERN
\cite{Scheurlein:2002} and our results.

\begin{figure}[htbp]
\centering
\includegraphics[width=0.6\textwidth,clip=]{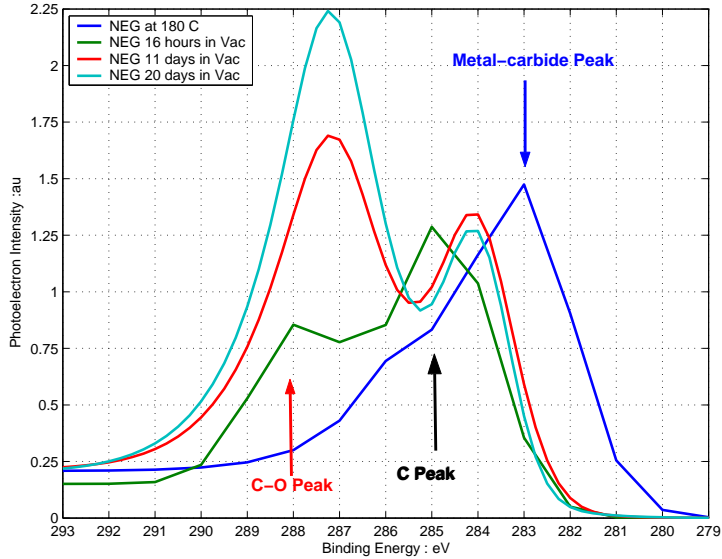}
\caption{XPS C peak of a TiZrV film after the end of activation}
\label{figXPSNEG3pk}
\end{figure}

\medskip
XPS analysis was carried out to observe the evolution of the
carbon chemistry during this 20 days and compared to the XPS
spectrum taken after the end of the activation. The XPS spectrum
shows a slight rise of the oxygen peak and a double valency carbon
1s peak. The elemental carbon peak has an energy of $\sim$285~eV,
the oxidized carbon peak has an energy of $\sim$288~eV, and the a
typical metal-carbide peak is around $\sim$283~eV. The oxidized
carbon peak after the end of the activation at 288~eV was barely
present, Fig.\ref{figXPSNEG3pk}, blue plot. Only The C peak and
Ti-C peak is present, the NEG being at 180$^\circ$C. A
cleanly-scraped carbon surface, measured at room temperature, will
only present a peak at 285~eV. After sixteen hours of pumping,
green plot, the oxidized state of the carbon shows up, and the
carbide peak is gone. This state, oxidized, becomes dominant after
eleven days of pumping and keep increasing after 20 days, gray
curve. In the last cases the NEG was at room temperature. The
plots presented in Fig.\ref{figXPSNEG3pk} are a fit of the actual
data. This increase of the C-O peak implies that the SEY should
also increase, as is usually the case for oxidized metal surfaces.

\medskip
The roughness (R) of the TiZrV film is unknown. This roughness
affects not only the SEY \cite{Hilleret} but also the pumping
capacity and speed of the NEG. TiZrV deposited on S/S is
relatively smooth (R$\simeq$1) and is much rougher on an aluminium
substrate \cite{benven:2001}. Let's assume that our system was for
twenty days at $\sim$3.10$^{-10}$~Torr, the CO being present at
15\% of the total spectrum, and that the sticking coefficient
$\sigma$ for CO is 0.4. The sticking coefficient decreases when
the TiZrV NEG has pumped almost half of a monolayer
($1~ML\sim$10$^{15}$~molecules-cm$^{-2}$ for a smooth surface) by
a factor 10 after reaching 1~ML, and by a factor 100 after 10~ML
\cite{benven:2001}. A monolayer of a surface (ML$_s$) can be
defined as~: $ML_s = ML \times R$ \cite{flp}. The frequency of
collision $\nu$ (molecules.s$^{-1}$.cm$^{-2}$)
equation~\ref{equfreqcollision} and the total amount of CO per
cm$^{-2}$ pumped by the NEG in one day is given by
equation~\ref{equCOpumped}.

\begin{minipage}[h]{.45\linewidth}
\begin{equation}
\nu = 3.5123 \ 10^{22} \frac{P}{\sqrt{M \ T}}
\label{equfreqcollision}
\end{equation}
\end{minipage}
\begin{minipage}[h]{.5\linewidth}
\begin{equation}
N_{CO} = \nu \times 86400 \times \sigma
\label{equCOpumped}
\end{equation}
\end{minipage}

where P is the pressure in Torr, M the atomic mass and T is the
temperature in K.

\medskip
After one day of pumping the NEG would have pumped $\sim$0.6~ML.
Assuming that for ten days the sticking coefficient is 0.04, the
NEG would have pumped an additional $\sim$0.6~ML. According to the
rough calculation for the total amount of CO pumped by the NEG,
the XPS of 11 days and 20 days should be similar, as the NEG is
basically saturated. This increase of the C-O peak,
Fig.\ref{figXPSNEG3pk}, suggests that the surface still had some
remaining pumping speed after eleven days of vacuum exposure.
Hence having a roughness R~$>$~1.
\newline NEG provides a nice solution for a distributed pumping
inside a vacuum chamber, when activated. Our small sample, a few
cm$^2$, being at 300~K pumped all of the residual gas in the
chamber except CH$_4$. In a few meters long NEG coated chamber the
contamination comes from outgassing surfaces outside this chamber,
since the NEG film is a diffusion barrier for outgassing molecules
of the substrate. Depending of the average vacuum in the machine
and the length of the NEG chambers, the recontamination of the
surface might take longer than for our sample. Hence, the SEY
might not increase as rapidly as measured here,
Fig.\ref{figTiZrV0-3000evA} and Fig.\ref{figTiZrV0-280evA}.

\medskip
Also, it has to be taken into account that the pumping lifetime of
thin film NEG depends on its thickness and the number of
activation cycles \cite{benven:2001}. The influence of the
thickness of the film should also be taken into account, when
calculating the impedance for the image current on the vacuum
chamber wall, due to a passing particle beam. The requirement on
the conductivity and the good mechanical performance after an
in-situ bake of the substrate is of importance. \newline For the
main damping rings, current design will use aluminium-alloy
chamber. Aluminium looses mechanical strength when heated above
150$^\circ$C. An adequate substrate for TiN or TiZrV which
fulfills the mechanical requirements is the alloy, Al~6061, or
Al~6060 which is easier to extrude. Their conductivity is
$\sim$1.4 times lower than for pure aluminium. SEY measurements on
a TiN film deposited on Al~6061 has already been carried out at
SLAC \cite{kirby:2000}.

\section{Conclusion}

We have presented a report on the status of the SEY experiment
carried out at SLAC. Description of our experimental system has
been presented.

\medskip
First results on as-received TiN sample and on an as-received
TiZrV getter have been shown. In the case of the getter, the
influence of the activation and recontamination by its pumping
action were investigated. The maximum SEY $\delta$ increased from
$\sim$1.3 to $\sim$1.5 after eleven days and to $\sim$1.6 after twenty-two
days of exposure to a vacuum of $\sim$3.10$^{-10}$~Torr. Our SEY
results seem to disagree with CERN
\cite{Scheurlein:2002}. First of all, our starting $\delta_{max}$
is 1.3 compare to 1.1 \cite{Scheurlein:2002}. Second, we have an
increase in $\delta_{max}$ above the CERN-predicted 1.35
\cite{Scheurlein:2002}. This matter should be investigated further
as the implication to electron cloud development is of importance
for a positively charged beam running in an accelerator.

\medskip
Values of $\delta$ for energies below 20~eV should be used
carefully if plugged into simulation. It is planned to investigate
further this part of the SEY curve, as has been done at CERN
\cite{cimino:2003}.

\medskip
Additionally, we will study the influence of various treatments,
such as the in-situ bakeout of TiN, electron conditioning applied
for the NLC case, and also the influence of ion conditioning. By
conditioning we mean bombarding the surface with a given spectrum
in energy of electrons and ions. Different species of ions can
also be investigated. \newline It has to be stressed that
conditioning (dose effect) is a very efficient way of lowering the
SEY of any technical surfaces to almost the same value ($\delta =
1.2$); and thus independently of the initial $\delta$
\cite{Hilleret}. However, such measurements are time consuming and
might not be relevant to the operation of an accelerator,
depending on the flux of electrons associated with
post-commissioning production operation.

\section{Acknowledgments}

We would like to thank P.~He and H.C.~Hseuh at BNL for providing
the TiN samples and the EST group from C.~Benvenuti at CERN for
the TiZrV sample. We also thank A.~Wolski at LBNL for shepherding
the production of sample plates, and in the near future, for thin
film samples coming from LBNL.  Most valuable was the work of G.
Collet and E. Garwin, SLAC, for converting and baking the XPS
system for use on SEY measurements.

\medskip


\end{document}